\documentclass[aps,prb,amsmath,amssymb,twocolumn,10pt,footinbib]{revtex4-1}
\usepackage{graphicx}
\usepackage{graphics,color}
\usepackage{hyperref}
\usepackage{hypernat}
\usepackage{simplewick}

\newcommand{\sms}{\;\;}
\newcommand{\ve}{\mathbf}

\newcommand{\D}{{\rm d}}

\begin{document}

\title{Bosonic field theory of tunable edge magnetism in graphene}
\author{Manuel J. Schmidt}
\affiliation{Institut f\"ur Theoretische Festk\"orperphysik, RWTH Aachen University, 52056 Aachen, Germany}
\affiliation{Department of Physics, University of Basel, Klingelbergstrasse 82, 4056 Basel, Switzerland}
\date{\today}

\begin{abstract}
A bosonic field theory is derived for the tunable edge magnetism at graphene zigzag edges. The derivation starts from an effective fermionic theory for the interacting graphene edge states, derived previously from a two-dimensional interacting tight-binding model for graphene. The essential feature of this effective model, which gives rise to the weak edge magnetism, is the momentum-dependent non-local electron-electron interaction. It is shown that this momentum-dependence may be treated by an extension of the bosonization technique, and leads to interactions of the bosonic fields. These interactions are reminiscent of a $\phi^4$ field theory. Focussing onto the regime close to the quantum phase transition between the ferromagnetic and the paramagnetic Luttinger liquid, a semiclassical interpretation of the interacting bosonic theory is given. Furthermore, it is argued that the universal critical behavior at the quantum phase transition between the paramagnetic and the ferromagnetic Luttinger liquid is governed by a small number of terms in this theory, which are accessible by quantum Monte-Carlo methods.
\end{abstract}

\maketitle

\section{Introduction}

Graphene is a two-dimensional hexagonal lattice of carbon atoms and shows many interesting features, ranging from Klein tunneling to anomalous quantum Hall behaviour.\cite{graphene_review_1} Also effects of electron-electron interactions in graphene have recently started to attract much interest. Because of the strong confinement of the electron wave functions to only one layer of carbon atoms, the interactions are very strong, as compared to other quasi two-dimensional systems. Yet, the large Fermi velocity and the vanishing density of states at the charge neutrality point efficiently suppress the manifestation of interaction effects in bulk graphene.

At graphene zigzag edges, however, states with a small kinetic energy, so called edge states, give rise to a strongly enhanced local density of states. These states are very susceptible to the electron-electron interactions: the kinetic energy of edge states is small so that the physics of the electrons in these exponentially localized states is dominated by electron-electron interaction. The latter drives the edges of graphene to a ferromagnetically ordered state, a phenomenon known as edge magnetism. This magnetic ordering has been predicted theoretically on the basis of many different methods for interacting electrons, such as mean-field approximations,\cite{edge_states_fujita_1996,jung_em_mean_field_2009} ab-initio calculations,\cite{son_abinitio_prl_2006,son_half_metallic_gnrs_2007} quantum Monte Carlo simulations,\cite{feldner_qmc_2010} and bosonization.\cite{bosonization_dmrg_hikihara_2003,tem_schmidt_loss_2010} Recently, also experimental evidence for edge magnetism was reported.\cite{joly_em_experimental1_2010,crommie_experimental_2011}

Usually, the edge states in graphene are considered to have negligibly small kinetic energies as compared to the typical bulk states. Indeed, edge states are exact zero energy modes of the pure nearest-neighbor $\pi$-band hopping Hamiltonian.\cite{edge_states_fujita_1996} In other words, the bandwidth of the edge states is zero in this simplest model. However, it has recently been shown that the edge state bandwidth, and therewith the Fermi velocity $v_F$ of these one-dimensional states, can be tuned over a wide range, and in many ways.\cite{soi_schmidt_loss_2010,tem_schmidt_loss_2010,ed_luitz_2011} A sufficiently large edge state bandwidth may reduce and even completely suppress the edge magnetism, so that the strength and also the type of the edge magnetism may be tuned experimentally.\cite{tem_schmidt_loss_2010}

In Ref. \onlinecite{tem_schmidt_loss_2010}, it was shown that (tunable) edge magnetism can be described on the basis of an effective one-dimensional edge state model which has been derived by projecting the Hubbard Hamiltonian of the two-dimensional honeycomb lattice onto the edge states. It was recently rigorously proven\cite{affleck} that the exact ground state of this model is ferromagnetic as long as the bandwidth is below a certain critical value.

In an exact diagonalization analysis of this edge state model,\cite{ed_luitz_2011} it was shown that the essential features of this model are (i) a momentum-dependent interaction vertex and (ii) the complete absence of umklapp scattering. Essentially, (ii) allows for the existence of ferromagnetism in one dimension, which is usually forbidden by the Lieb-Mattis theorem.\cite{lieb_mattis} However, (ii) alone only allows a 1st order phase transition from a paramagnetic Luttinger liquid, appearing in the limit of large edge state bandwidths, to a maximally polarized state as for smaller bandwidths. This happens because, once the system is beyond the Stoner point ($v_F\lesssim U/\pi$, where $v_F$ is the Fermi velocity and $U$ is the Hubbard interaction on the honeycomb lattice), the interaction energy gain associated with increasing the spin-polarization is always higher than the corresponding kinetic energy penalty, independently of the actual spin-polarization. Thus, the system with feature (ii) alone flows to full polarization immediately as it hits the Stoner instability. The momentum-dependence of the interactions (i) provides a mechanism to stop this flow: essentially the interaction strength is inversely proportional to the spin-polarization, so that the interaction energy gain becomes smaller as the spin-polarization is increased. Thus, the flow to higher polarizations is stopped at a certain spin-polarization and the system is stabilized in an itinerant weak ferromagnetic state.\footnote{For a detailed discussion of this mechanism, see Ref. \onlinecite{ed_luitz_2011}.}

On the basis of the identification of these two important edge state features (i) and (ii), a generalized edge state model was proposed and analyzed by exact diagonalization in Ref. \onlinecite{ed_luitz_2011}. In this paper, an interacting bosonic field theory is derived from this generalized fermionic edge state model, thus overcoming the restriction to small systems of the exact diagonalization analysis. For this the well-known bosonization technique\cite{giamarchi_book} is adapted to the momentum-dependent interactions. The proper treatment of the momentum-dependence of the interaction vertex is essential for the analysis of the weak edge magnetism. The resulting bosonic action contains interactions of the boson fields, giving rise to an appealing interpretation as a generalized Landau functional of the form $a m^2 + b m^4 + ...$, with $m$ the local magnetization. The actual bosonic theory seems to be very complicated, but it will be argued that only few terms contribute to the critical properties at the transition between the ferromagnetic and the paramagnetic Luttinger liquid. The critical theory, which is proposed for this transition, is a 1+1 dimensional classical field theory with a real action. As such, this critical bosonic theory may be simulated with Monte-Carlo methods and does not suffer from the infamous sign problem.

The paper is organized as follows. In Sec. \ref{sect_model}, the generalized edge state model and a simple fermionic mean-field analysis of it is discussed, in order to gain an overview of the phase diagram and the relevant mechanisms. Section \ref{sect_bosonization} contains the derivation of the bosonic theory, which is then analyzed in Sec. \ref{sect_analysis}. A summary and a critical discussion of the results may be found in Sec. \ref{sect_discussion}.

\section{The fermionic model\label{sect_model}}
\begin{figure}[!ht]
\centering
\includegraphics[width=245pt]{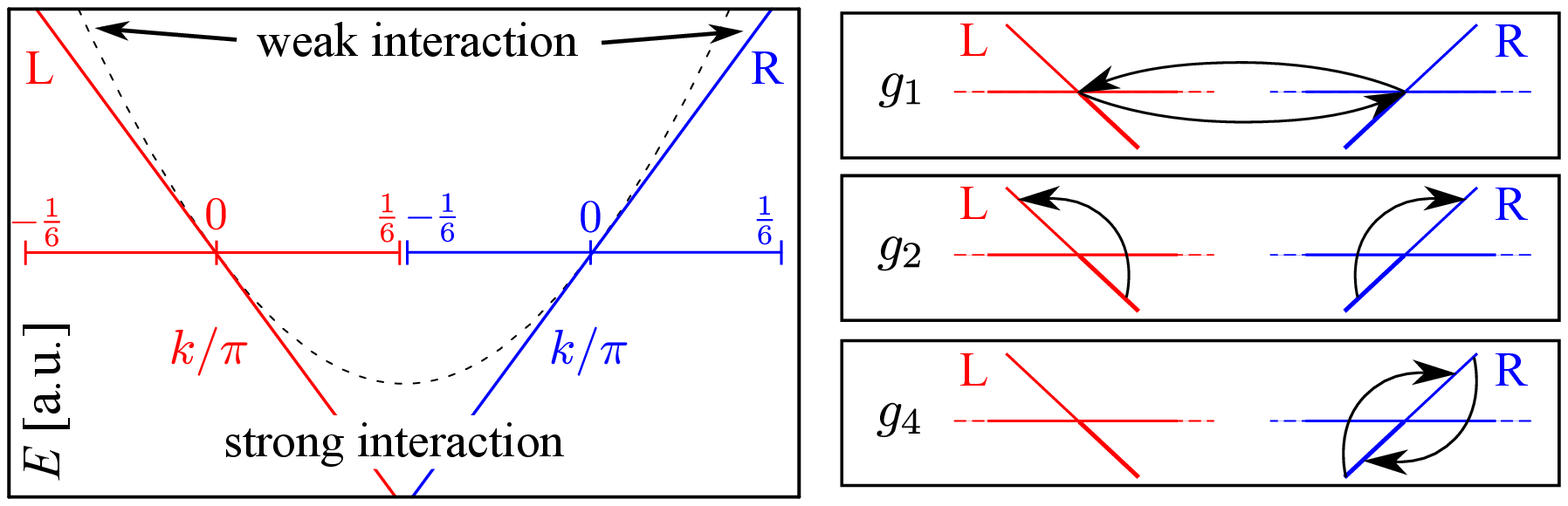}
\caption{(Color online) Left: The linearized dispersion, described by $H_0$. The dashed line indicates the original cosine dispersion of direct model derived in Ref. \onlinecite{tem_schmidt_loss_2010}. For $\Gamma_1>0$, the electrons at high energy are weakly interacting while the low-energy sector is strongly interacting. The momenta of the left- and right-movers are centered around their respective Fermi momenta. Right: The allowed interaction processes $g_1$ (backscattering), $g_2,g_4$ (forward scattering). Umklapp processes are forbidden for the edge states (see text).}
\label{fig_fermionic_model}
\end{figure}

The derivation of the bosonic field theory starts from the generalized edge state model, introduced in Ref. \onlinecite{ed_luitz_2011}. The umklapp scattering is not allowed for the edge states considered here, so that the Hamiltonian consists of three terms
\begin{equation}
H = H_0 + H_1^{\rm fs} + H_1^{\rm bs},
\end{equation}
describing the kinetic energy, the forward scattering and the backscattering, respectively. The linearized kinetic energy reads
\begin{equation}
H_0 = v_F \sum_{\substack{r=R,L\\\sigma=\uparrow,\downarrow}} \sum_{k=-\pi/6}^{\pi/6} (r k) c_{kr\sigma}^\dagger c_{kr\sigma},
\end{equation}
where $v_F$ is the Fermi velocity, $k$ is the momentum along the zigzag edge in units of the graphene lattice constant $a\simeq 2.4$\AA, $r=R,L$ stands for right- and left-moving electrons, respectively, and $c_{kr\sigma}$ annihilates an $r$-moving spin-$\sigma$ electron with momentum $k$. When used in an equation, $r=\pm1$ for $R,L$, respectively. Note that $k=0$ is defined separately for left- and right-movers (see Fig. \ref{fig_fermionic_model}). The domain of the edge states ($[-\frac\pi6,\frac\pi6]$ for each $r=R,L$) is restricted to one third of the total Brillouin zone $[-\pi,\pi]$ (see Ref. \onlinecite{ed_luitz_2011}). For convenience the Fermi level is set to zero energy, i.e. $k_F=0$ for left- and right-movers. Note that, due to the absence of umklapp scattering, this choice can be made without loss of generality.

The interaction Hamiltonian $H_1 = H_1^{\rm fs} + H_1^{\rm bs}$ is most conveniently defined in $k$-space. The forward scattering Hamiltonian, corresponding to $g_2$ and $g_4$ processes\cite{giamarchi_book} (see Fig. \ref{fig_fermionic_model}) reads
\begin{multline}
H^{\rm fs}_1 = \frac UL \sum_{r,r'} \sum_{k,k',q} S^r_{k+q} S^r_k S^{r'}_{k'-q} S^{r'}_{k'}\times \\ :c^\dagger_{k+qr\uparrow} c_{kr\uparrow} c^\dagger_{k'-q r'\downarrow} c_{k'r'\downarrow}:,\label{def_h_fs},
\end{multline}
where the $k$-space summations are restricted such that all momentum arguments of the electron operators are in the interval $[-\frac\pi6,\frac\pi6]$. $L$ is the length of the edge in units of the lattice constant $a$ and the total strength of the interaction is expressed by a Hubbard $U$. The colons indicate normal ordering of the fermion operators with respect to the non-interacting Fermi sea. The factors
\begin{equation}
S_k^r = \sqrt{1-r\Gamma_1 k}
\end{equation}
with $\Gamma_1\in[0,\frac6\pi]$ parametrize the momentum-dependence. For $\Gamma_1=0$, the interaction vertex is momentum-independent. This limit corresponds to the forward scattering in a usual one-dimensional Hubbard chain.\cite{ed_luitz_2011} $\Gamma_1 = \frac6\pi$ corresponds to the interaction vertex of edge states. Although this work is finally targeted at the edge state limit $\Gamma_1=\frac6\pi$, it is convenient for the bosonization to keep $\Gamma_1$ as a free parameter.

The backscattering Hamiltonian, describing $g_1$ processes (see Fig. \ref{fig_fermionic_model}), is given by
\begin{multline}
H^{\rm bs}_1 = \lambda_{\rm bs} \frac{U}L \sum_r \sum_{k,k',q} S^r_{k+q} S^{-r}_k S^{-r}_{k'-q} S^r_{k'} \times\\c^\dagger_{k+q,r,\uparrow} c_{k,-r,\uparrow} c^\dagger_{k'-q,-r,\downarrow} c_{k',r,\downarrow},\label{eq_ham_backscattering}
\end{multline}
where the additional parameter $\lambda_{\rm bs}$ has been introduced in order to be able to tune the backscattering strength. $\lambda_{\rm bs}=1$ corresponds to the physical SU(2) invariance, which is obtained by a direct derivation from the two-dimensional Hubbard model.\cite{tem_schmidt_loss_2010}

The factors $S^r_k$ enter $H_1^{\rm bs}$ in a similar manner as they appear in $H_1^{\rm fs}$. In both cases, for each electron operator participating in the interaction, the factors $S^r_k$ rescale the interaction strength: high-energy electrons suppress the interaction and low-energy electrons increase it. This means that exactly those processes which usually invalidate the bosonization technique beyond a certain interaction strength, namely the ones involving electrons far from the Fermi level, are suppressed.\footnote{Note that the processes involving electrons deep within the Fermi sea are enhanced. It turns out, however, that this is overcompensated by the suppression at high energies.} Indeed, it turns out that a proper treatment of the $S_k^r$ factors in the bosonization results in a well-controlled theory also beyond the Stoner point, i.e. for strong interactions.

\begin{figure}[!ht]
\centering
\includegraphics[width=160pt]{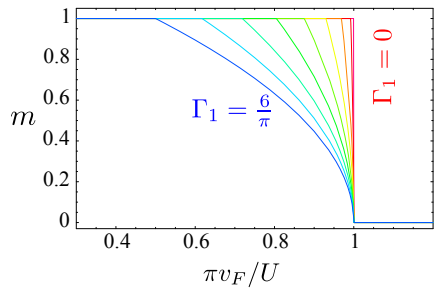}
\caption{(Color online) Spin-polarization $m$ from the fermionic mean-field theory. The different $m$ curves correspond to different momentum dependencies $\Gamma_1=0,...,\frac6\pi$.}
\label{fig_mf_phase_diagram}
\end{figure}

In order to demonstrate the significance of $\Gamma_1$, I proceed with a simple fermionic mean-field (fMF) analysis of $H_0+H_1^{\rm fs}$.\footnote{See also Ref. \onlinecite{ed_luitz_2011}.} The parameter $\Delta k$ describes the splitting of the Fermi points $k_{Fr\sigma}=r\sigma\Delta k$ of $r$-moving spin-$\sigma$ electrons ($\sigma=\pm1$ for $\uparrow$- and $\downarrow$-spins, respectively, when used in equations). Obviously, $\Delta k$ is proportional to the spin-polarization $m = \frac6\pi \Delta k\in[-1,1]$. The variational energy as a function of the spin-polarization $m$, corresponding to a fMF treatment, is
\begin{equation}
E_{\rm fMF}(m) =  \frac{\pi v_F - U}{36} m^2 + \frac{U \Gamma_1^2 \pi^2}{5184} m^4.
\end{equation}
For $\pi v_F>U$, $E_{\rm fMF}(m)$ is minimal for zero polarization. This is the paramagnetic Luttinger liquid phase for weak interactions. For $\pi v_F<U$, however, the ground state has a non-zero spin-polarization $m\neq 0$. Fig. \ref{fig_mf_phase_diagram} shows $m$ as a function of the Fermi velocity $v_F$.

The size of the spin-polarization is controlled by the $m^4$ term, which is proportional to $\Gamma_1^2$. For $\Gamma_1=0$, the magnetization instantly jumps from zero to its maximum value as the fMF Stoner point $\pi v_F=U$ is crossed. In the $m=1$ state all up-spin edge states are occupied and all down-spin edge states are empty. Such a scenario where the band edges determine the physics can evidently not be treated within bosonization. Fortunately, the $\Gamma_1>0$ in actual graphene zigzag edge states gives rise to a controlled intermediate regime of weak itinerant ferromagnetism ($0<|m|<1$), in which bosonization is applicable.

Compared to the fMF prediction for the position of the Stoner instability, exact diagonalization gives a transition shifted to higher $U$.\cite{ed_luitz_2011} A previous analysis, based on a combination of fMF and naive bosonization,\cite{tem_schmidt_loss_2010} failed to explain this shift as this analysis was valid only sufficiently far from the Stoner point. One goal of the extended bosonization, presented in the following, is to overcome this limitation.

Note that in Ref. \onlinecite{affleck} the critical properties at the ferromagnetic transition differ from the one discussed here. This due to different assumptions about the filling as the kinetic energy ($v_F$ in this work and $\Delta$ in Ref. \onlinecite{affleck}) is changed; in the present work, a constant number of electrons in the edge states is assumed, while Ref. \onlinecite{affleck} assumes that the total filling approaches a trivial limit (completely filled or completely empty edge states) as the system becomes non-magnetic.

\section{Bosonization of the edge state model\label{sect_bosonization}}

In its usual form,\cite{giamarchi_book} bosonization maps an interacting one-dimensional fermionic model onto an essentially non-interacting bosonic model.\footnote{The sine-Gordon term, resulting from the backscattering processes, is well studied and can be analyzed by RG methods (see Ref. \onlinecite{giamarchi_book}). Its proper treatment results in a non-interacting bosonic field theory with renormalized parameters.} However, two prerequisites are essential for this mapping: (i) the single particle spectrum must be linear, and (ii) the interaction vertex of a process $g_i$ (see Fig. \ref{fig_fermionic_model}) must be a constant. For the edge states considered here, (i) is a good approximation, as has been shown in Ref. \onlinecite{ed_luitz_2011}. However, the momentum dependence of the interaction, which conflicts with (ii), is an essential feature of edge states, responsible for the stabilization of the weak ferromagnetism. Thus, (ii) is a priori not valid and the usual bosonization technique must be adapted in order to be applicable to the interactions of the form (\ref{def_h_fs}).

Before discussing the momentum-dependent bosonization, it is instructive to revisit the usual bosonization and its breakdown at the Stoner instability.

\subsection{Naive bosonization}
If the interaction-induced electron-hole excitations are restricted to a small region around the Fermi surface in which the interaction vertex does not vary much, the latter may be approximated by a constant, i.e., $S^r_k\simeq 1$. Reducing the functional form of an interaction vertex to constants is the central idea of g-ology, which is well justified for small interactions $U$. The resulting bosonic theory $H_s+H_c + H_1^{\rm bs}$ describes an ordinary spinful Luttinger liquid, with
\begin{equation}
H_\nu = \frac{1}{2\pi} \int\D x \left[ u_\nu K_\nu (\partial_x \theta_\nu)^2 + \frac{u_\nu}{K_\nu} (\partial_x \phi_\nu)^2\right] \label{boson_hamiltonian_0}
\end{equation}
and $u_\nu K_\nu = v_F$, $u_\nu/K_\nu = v_F (1\pm U/\pi v_F)$ for $\nu=c,s$, respectively. The bosonic field operators $\phi_\nu$ and $\theta_\nu$ obey the commutation relations $\left[\phi_\nu(x'),\partial_x\theta_{\nu'}(x)\right] = i\pi \delta_{\nu\nu'}\delta(x-x')$ and their mapping to the fermion operators $c_{kr\sigma}$ is given in Appendix \ref{appendix_bosonization_details}. Their physical interpretation becomes obvious by reexpressing the spatial derivatives of the field operators in terms of fermionic field operators $\hat\psi_{r\sigma}(x)$
\begin{align}
\pi \partial_x \phi_c(x) &= \sum_{r\sigma} \hat\psi^\dagger_{r\sigma}(x) \hat\psi_{r\sigma}(x)\label{bos_fields_1}\\
\pi \partial_x \theta_c(x) &= \sum_{r\sigma} r \hat\psi^\dagger_{r\sigma}(x) \hat\psi_{r\sigma}(x)\\
\pi \partial_x \phi_s(x) &= \sum_{r\sigma} \sigma \hat\psi^\dagger_{r\sigma}(x) \hat\psi_{r\sigma}(x)\\
\pi \partial_x \theta_s(x) &= \sum_{r\sigma} r \sigma \hat\psi^\dagger_{r\sigma}(x) \hat\psi_{r\sigma}(x).\label{bos_fields_2}
\end{align}
Obviously, the derivatives of the bosonic fields correspond to densities ($\phi$) and currents ($\theta)$ of charge ($\nu=c$) and spin $(\nu=s$).

The backscattering Hamiltonian $H_1^{\rm bs}$ [see Eq. (\ref{eq_ham_backscattering})] translates, for $S_k^r\simeq 1$, to a bosonic sine-Gordon term $[2\lambda_{\rm bs}U/(2\pi\eta)^2] \int\D x \cos(\sqrt8 \phi_s)$ which is responsible for restoring SU(2) invariance in the Luttinger-liquid regime\cite{giamarchi_book}. The exact diagonalization analysis in Ref. \onlinecite{ed_luitz_2011} suggests that, apart from restoring SU(2) symmetry, the backscattering is of minor importance in the magnetic phase. However, it cannot be excluded that $H^{\rm bs}_1$ changes the critical properties at the phase transition. Nevertheless, the general idea of the bosonic theory can be understood on the basis of the forward scattering interaction only. Therefore, the breakdown of the naive bosonic theory [Eq. (\ref{boson_hamiltonian_0})] is discussed on the basis of the forward scattering only.

This naive bosonic theory breaks down for $U>U_{\rm crit.}=\pi v_F$ where the term in the Hamiltonian (\ref{boson_hamiltonian_0}), controlling spin density fluctuations $(u_s/K_s) \phi_s'^2$, becomes negative. The reason for this breakdown is that the particle-hole excitations are not restricted to small energies for such strong $U$. In a sense, an interaction $U>U_{\rm crit.}$ drives the excitations to higher and higher energies, i.e. to regions far from the Fermi level (compare also the fMF analysis in Sec. \ref{sect_model}). As this interaction-driven flow to high energies is not limited in one dimension, the particle-hole excitations hit the band edge. This regime in which the band edge becomes important cannot be described by bosonization. To see the breakdown on a purely formal level it is sufficient to note that for $U>U_{\rm crit.}$ the bosonic Hamiltonian is not bounded from below.

For actual edge states, however, the interaction becomes effectively weaker at higher energies because of the factors $S_k^r$. Due to this suppression of the interaction, the flow to higher energies is stopped before it hits the band edge, so that this regime can be described by bosonization in principle. However, since the approximation of constant $S_k^r$ is not valid, the usual bosonization technique must be extended in order to be able to account for the $k$-dependence of the interaction vertex.

As described in Ref. \onlinecite{tem_schmidt_loss_2010}, it is possible to go beyond the critical interaction $U_{\rm crit.}$ by a combination of fermionic mean-field theory and bosonization: the mean-field solution predicts a splitting between the up-spin and the down-spin Fermi level, which, due to the momentum-dependence of the interaction, effectively reduces the interaction strength for electrons at these new (spin-dependent) Fermi levels. Thus, if the bosonization is performed on the basis of the fermionic mean-field theory, it turns out to be stable also for $U>U_{\rm crit.}$. In a sense, the interaction is separated into strong parts which may be treated within mean-field theory, and weak parts that are accessible via the usual bosonization technique.

This combined method, used in Ref. \onlinecite{tem_schmidt_loss_2010}, has severe limitations. Most importantly, it is only valid sufficiently far from the critical point. Also, the approximation of constant interaction vertices within the bosonization part of the method is not controlled. Thus, a more sophisticated approach is used in this work, treating the full momentum dependence within the bosonization technique without resorting to mean-field theory. It is shown that the momentum-dependence of the fermionic interactions translate to interactions of the bosonic fields.

\subsection{Boson interactions from forward scattering}
Usually, interaction terms of the forward scattering form $\hat\psi^\dagger_{r\uparrow} \hat\psi_{r\uparrow} \hat\psi_{r'\downarrow}^\dagger\hat\psi_{r'\downarrow}$ may be bosonized by means of the bosonization identity $\hat\psi^\dagger_{r\sigma} \hat\psi_{r\sigma} = \frac r\pi \partial_x\phi_{r\sigma}$. The Fourier transform of $H_1^{\rm fs}$ to real-space, however, reads
\begin{multline}
H_1^{\rm fs} = U\sum_{rr'}\int\D x \\ \left[S^r(-\hat k)\hat\psi^\dagger_{r\uparrow}\right]\left[S^r(\hat k)\hat\psi_{r\uparrow}\right]\left[S^{r'}(-\hat k)\hat\psi^\dagger_{r'\downarrow}\right]\left[S^{r'}(\hat k)\hat\psi_{r'\downarrow}\right],\label{fs_int_momentum_dep}
\end{multline}
with $S^r(\pm \hat k) = \sqrt{1 \mp r\Gamma_1 \hat k}$ and $\hat k = -i\partial_x$ the momentum operator. Eq. (\ref{fs_int_momentum_dep}) may be expanded in powers of $\Gamma_1$ so that a generalized bosonization identity for
\begin{equation}
\left[(i\partial_x)^n\hat\psi^\dagger_{r\sigma}\right]\left[(-i\partial_x)^m\hat\psi_{r\sigma}\right]\label{ferm_int_form}
\end{equation}
for general integers $n,m$ is needed. This is derived in Appendix \ref{appendix_bosonization_details}. Essentially the bosonic version of Eq. (\ref{ferm_int_form}) gives rise to a $(n+m+1)$th power of first derivatives of the boson fields, so that Eq. (\ref{fs_int_momentum_dep}) leads to interactions in the bosonized theory.

In principle, the bosonic momentum-dependent forward scattering has infinitely many terms. However, almost all terms involve at least second order spatial derivatives of the fields. These higher field derivatives correspond to spatial fluctuations of densities and currents. Their size can be estimated for the largest allowed momenta; it turns out that these terms are small. Furthermore, one is foremost interested in the first order derivatives, since these are directly connected to the global order parameters. Most importantly, the field $\phi_s$ is related to the local spin-polarization $m$
\begin{equation}
\left<\phi_s'\right> = \frac{\pi}{3\sqrt2} m,
\end{equation}
where $|m|=1$ for maximal polarization. In order to see the structure of the bosonic theory more clearly it is convenient to partition $H_1^{\rm fs}=H_1^{{\rm fs},(0)} +H_1^{{\rm fs},(1)}$, with $H_1^{{\rm fs},(0)}$ containing only terms involving the first derivatives of the boson fields, while all terms containing higher order derivatives are collected in $H_1^{{\rm fs},(1)}$. 

The details of the momentum-dependent bosonization procedure can be found in Appendix \ref{appendix_bosonization_details}. Interestingly, it turns out that all orders higher than $\Gamma_1^2$ in $H_1^{{\rm fs},(0)}$ cancel exactly, so that
\begin{widetext}
\begin{multline}
H_1^{{\rm fs},(0)} = \frac1{2\pi}\int \D x \biggl\{ \frac{U}{\pi}\left[\phi_c'^2 - \phi_s'^2\right] + \frac{U\Gamma_1}{\sqrt2\pi} \left[2 \phi_s' \theta_s'\theta_c' + \phi_s'^2 \phi_c' - \theta_c'^2 \phi_c' - \theta_s'^2\phi_c' - \phi_c'^3  \right] \\ +  \frac{U\Gamma_1^2}{8\pi} \biggl[\theta_c'^4+\theta_s'^4+\phi_c'^4+\phi_s'^4
- 2\theta_c'^2\theta_s'^2 - 2\phi_c'^2\phi_s'^2+ 2\theta_c'^2\phi_c'^2 + 2\theta_s'^2\phi_c'^2 + 2\theta_c'^2\phi_s'^2 + 2\theta_s'^2\phi_s'^2  -8\theta_c'\theta_s'\phi_s'\phi_c'\biggr]\biggr\} . \label{fws0_ham_bos}
\end{multline}
\end{widetext}
Eq. (\ref{fws0_ham_bos}) is one of the main results of this paper. It shows how the strong momentum dependence of the interaction vertex ($\Gamma_1\neq 0$) gives rise to interactions in the bosonic theory.

The infinitely many terms in $H_1^{{\rm fs},(1)}$ cannot be written down in a closed form. Fortunately, it turns out that this is not needed for the further analysis. For the leading order in $\Gamma_1$ one finds
\begin{equation}
H_1^{{\rm fs},(1)} = \frac1{2\pi}\int\D x\frac{U\Gamma_1^2}{4\pi} \left[\phi_s''^2 - \phi_c''^2\right] + O(\Gamma_1^3).\label{bos_fw_1}
\end{equation}
The higher orders in $\Gamma_1$ also involve higher orders in $k$. As there was no sign of short wavelength instabilities in the exact diagonalization study of this model\cite{ed_luitz_2011}, one may safely restrict $k\ll\Gamma_1^{-1}$ and drop the $O(\Gamma_1^3)$ terms.

On the first sight, the negative sign in front of the $\phi''^2_c$ term appears to lead to a Hamiltonian which is not bounded from below. However, in practice, as $k$ cannot be larger
than $\Gamma_1^{-1}$ and all $k$ integrals are regularized by this UV cutoff, this term is always smaller than the $\phi'^2_c$ term in $H_1^{{\rm fs},(0)}$. Also, it creates no mexican-hat minimum at finite $k$. Thus, the $\phi''^2_c$ term in $H_1^{{\rm fs},(1)}$ may safely be ignored.

In the critical theory $H_1^{{\rm fs},(1)}$ is needed to control the short wavelength fluctuations in the correlation function $\left<\phi'_s\phi'_s\right>$ at the critical point. This is illustrated on the basis of a simplified classical model for the $\phi_s$ field. Consider the Hamiltonian density
\begin{equation}
\mathcal H = \tau \phi_s'^2 + \alpha \phi_s''^2 + \phi_s'^4,\label{class_model}
\end{equation}
where $\phi_s(x)$ is assumed to be a real classical field in this paragraph (and only in this paragraph). At the mean-field critical point of $\mathcal H$, the prefactor $\tau$ of the $\phi'^2$ term becomes zero. In the Landau mean-field theory, where the interaction $\phi'^4$ is neglected, the term $\alpha\phi_s''^2$ is then the only non-zero term at the critical point, controlling the behavior of the correlation function
\begin{equation}
\left<\phi_s'(x)\phi_s'(0)-\phi_s'(0)\phi_s'(0)\right>^{\rm class.}_{\mathcal H} \propto \int \D k \frac{k^2 (e^{ikx}-1)}{\tau k^2+ \alpha k^4}.\label{simplified_corr}
\end{equation}
For $\tau=0$, the correlation function (\ref{simplified_corr}) is proportional to $-y/\alpha$. Note the similarity of (\ref{simplified_corr}) to the correlation function of a standard $\phi^4$ model. This is a consequence of the assumption of a classical field in $\mathcal H$. Of course, the $\phi^4$ model in one dimension does not give rise to a thermodynamic phase transition. Treating $\phi_s$ correctly as a bosonic field conjugate to $\theta_s$ leads to a theory that is structurally different from a classical $\phi^4$ model; because of the anisotropy in time and space it is not possible to reduce the orders of the spatial derivatives by one, as in the classical model (\ref{class_model}).

 The higher order terms in Eq. (\ref{bos_fw_1}) involve higher order spatial derivatives, adding $k^6$, $k^8$,... terms to the denominator in (\ref{simplified_corr}). These do not change the infrared singularities and are thus, in analogy to the usual argumentation in the theory of critical phenomena,\cite{kleinert_book} expected to be irrelevant for the critical behavior.

Note that $\phi_s''^2$ is not the only term which controls the field fluctuations at the critical point. The sine-Gordon term derived in the next subsection has a similar effect.

\subsection{Boson interactions from backscattering}
In this section, the backscattering processes are bosonized. For a constant interaction vertex, backscattering gives rise to a sine-Gordon Hamiltonian for the boson fields. Here, the momentum-dependence of the interaction vertex is taken into account, which leads to a generalized sine-Gordon Hamiltonian. The Fourier transform of the backscattering Hamiltonian [Eq. (\ref{eq_ham_backscattering})] to real space reads
\begin{multline}
H^{\rm bs}_1 = \lambda_{\rm bs} U \sum_{r} \int \D x \\ \left[S^r(-\hat k)\hat\psi^\dagger_{r\uparrow}\right]\left[S^{-r}(\hat k)\hat\psi_{-r\uparrow}\right]\left[S^{-r}(-\hat k)\hat\psi^\dagger_{-r\downarrow}\right]\left[S^{r}(\hat k)\hat\psi_{r\downarrow}\right].\label{bs_ham_real_space}
\end{multline}
The bosonization of (\ref{bs_ham_real_space}) proceeds along similar lines as the bosonization of the forward scattering Hamiltonian: the $S^r$ factors are expanded in powers of $\Gamma_1$ and each term is bosonized separately. Unfortunately, however, there is no cancellation of higher $\Gamma_1$ orders as in the forward scattering case. In exchange, a closed formula for the bosonic Hamiltonian of the momentum-dependent backscattering process can be given
\begin{widetext}
\begin{multline}
H_1^{\rm bs} = \frac{U}{(2\pi\eta)^2} \sum_r \int\D x \sum_{n_1,n_2,n_3,n_4}c_{n_1}c_{n_2}c_{n_3}c_{n_4} \\ \left[(ir\Gamma_1 \partial_x)^{n_1} e^{-2i\phi_{r\uparrow}}\right]\left[(ir\Gamma_1 \partial_x)^{n_2} e^{2i\phi_{-r\uparrow}}\right]\left[(-ir\Gamma_1 \partial_x)^{n_3} e^{-2i\phi_{-r\downarrow}}\right]\left[(-ir\Gamma_1 \partial_x)^{n_4} e^{2i\phi_{r\downarrow}}\right]\label{backscattering_full_bosonic_hamiltonian}
\end{multline}
\end{widetext}
where $c_n$ is the expansion coefficient of the power series of $\sqrt{1-x}=\sum_n c_n x^n$. The bosonic fields $\phi_{r\sigma}$ are commuting combinations of the four bosonic fields introduced in Eqs. (\ref{bos_fields_1}) - (\ref{bos_fields_2})
\begin{equation}
\phi_{r\sigma} = \frac1{\sqrt8} (r\phi_c + r\sigma \phi_s - \theta_c - \sigma\theta_s).
\end{equation}
For more details see also Appendix \ref{appendix_bosonization_details}. 

From Eq. (\ref{backscattering_full_bosonic_hamiltonian}) all terms of the bosonic backscattering can be written down systematically. It is instructive to arrange the terms in powers of $\Gamma_1$ and to reinstate the original bosonic fields $\phi_{c,s}$ and $\theta_{c,s}$. Up to fourth order in $\Gamma_1$, one finds
\begin{widetext}
\begin{equation}
H_1^{\rm bs} = \frac{2U}{(2\pi\eta)^2} \int\D x \left[1-\sqrt2\Gamma_1 \phi_c' +\frac{\Gamma_1^2}2(\phi_c'^2 + \phi_s'^2 - \theta_c'^2 - \theta_s'^2 )\right] \cos\sqrt8\phi_s + O(\Gamma_1^4). \label{bosonic_backscattering_red}
\end{equation}
\end{widetext}
Note that the terms of order $\Gamma_1^3$ vanish exactly in Eq. (\ref{bosonic_backscattering_red}). The terms of order $\Gamma_1^4$ and higher contain higher spatial derivatives of the bosonic fields. At the critical point, they are not needed to control the fluctuations of the field as this is already done by lower order terms. Thus, according to the arguments given at the end of the last subsection, these are expected to be small and irrelevant for the critical behavior.

\section{Qualitative analysis of the bosonic field theory\label{sect_analysis}}

In this section, an interpretation of the bosonic theory is presented. The goal of this section is to convey an impression of the significance of the individual terms for the ferromagnetic phase diagram. It should be emphasized that the following handwaving arguments are by no means exact. The actual solution of the full interacting field theory requires sophisticated methods and is beyond the scope of this work.

\subsection{Classical interpretation}
As already shown in Ref. \onlinecite{ed_luitz_2011}, the backscattering has only little impact on the phase diagram, so that one may start by dropping $H_1^{\rm bs}$. In this subsection, the classical vacuum of the energy functional $E_C[\phi_s,\phi_c,\theta_s,\theta_c]\equiv H_0 + H_1^{\rm fs}$ shall be found, assuming all fields to be classical (i.e. $[\phi_\nu,\theta_\nu]=0$). Only spatial derivatives of the fields enter the energy functional. Since the leading terms for the three fields $\phi_c,\theta_s,\theta_c$ have a positive coefficient, these fields are constant in the classical minimum of $E_C$ and drop out of the further analysis. Thus, it is sufficient to consider
\begin{equation}
\tilde E_C[\phi_s] = \frac1{2\pi} \left[\left(v_F - \frac U\pi\right) \phi_s'^2 + \frac{U\Gamma_1^2}{8\pi} \phi_s'^4 +\frac{U\Gamma_1^2}{4\pi} \phi_s''^2 \right] .
\end{equation}
Because of the positive sign in front of the $\phi_s''^2$ term, the optimal classical field is a linear function of $x$, i.e. $\phi_s(x) = \frac\pi{3\sqrt2} m x$, with $m$ the magnetization (cf. the fermionic mean-field), so that $\tilde E_C[\phi_s]$ reduces to a Landau function for $m$
\begin{equation}
\mathcal L[m] \propto \frac{\pi v_F - U}{36} m^2 + \frac{U \Gamma_1^2 \pi^2}{5184} m^4
\end{equation}
which {\it exactly} reproduces the fermionic mean-field theory of the generalized edge-state model. Of course, the actual theory is more complicated because of the quantum nature of the fields. Thus, the features which are not correctly predicted by the fermionic mean-field theory must be a consequence of the quantum nature of the bosonic theory. For instance, the classical treatment is not able to explain the dependence of the critical point on $\Gamma_1$ which was found in the exact diagonalization treatment.

\subsection{Quantum fluctuations around the saddle point solution}

In this subsection, the quantum nature of the bosonic fields is taken into account on the simplest level: by a variational technique, the interacting Hamiltonian is approximated by the 'closest' non-interacting Hamiltonian. This treatment is equivalent to a bosonic mean-field theory (bMF) and to a treatment of the quadratic fluctuation around the saddle point solution. Before this is done for the complicated Hamiltonian $H_0 + H_1^{\rm fs}$, a much simplified model of only one pair of bosonic fields $\phi,\theta$ with $\left[\phi(x'),\partial_x\theta(x)\right] = i\pi\delta(x-x')$ and a single $\phi'^4$ interaction is discussed. Consider the Hamiltonian density
\begin{equation}
\mathcal H^{(0)} = u K \theta'^2 + \frac uK\phi'^2 + \alpha \phi'^4,\label{toy_h0}
\end{equation}
with $u K = 1$ and $\frac uK = 1-U/\pi v_F \equiv \tau$. Up to a global prefactor, Eq. (\ref{toy_h0}) contains the spin sector terms of order $\Gamma_1^0$ and one of the $\Gamma_1^2$ terms, i.e. $\alpha\phi'^4$ from $H_1^{\rm fs,(0)}$. The $\phi'^4$ term has been chosen because $\tau$ is allowed to change its sign. Then the $\phi'^4$ bounds the Hamiltonian from below. Following the prescription in Appendix \ref{appendix_bMF}, the bMF approximation of Eq. (\ref{toy_h0}) is given by
\begin{equation}
\mathcal H^{(0)}_{\rm bMF} = \theta'^2 + \left[\tau + 6 \alpha \left<\phi'^2\right>\right]\phi'^2, \label{toy_bMF}
\end{equation}
so that one may introduce renormalized parameters $u^*$ and $K^*$ with
\begin{align}
u^*K^* &= 1 & \tau^* \equiv \frac{u^*}{K^*} &= \tau +6 \alpha \left<\phi'^2\right>.\label{toy_sc}
\end{align}

At this point the mechanism which is responsible for the dependence of the critical point on $\Gamma_1$ becomes obvious: compared to the fermionic mean-field theory, which is critical at $\tau=0$, the bosonic mean-field theory is critical at $\tau = -6\alpha\left<\phi'^2\right>$, with $\alpha\propto \Gamma_1^2$. In the language of the renormalization group, $\tau \rightarrow\tau^*$ corresponds to a mass renormalization, which is known to be non-universal, i.e., it usually depends on microscopic details. Here, these microscopic details enter the renormalized $\tau^*$ via the correlation function $\left<\phi'^2\right>$ in which the fields have the same spatial and temporal arguments. On a formal level, a clear sign for quantities which depend on microscopic details is the appearance of the high energy cutoff $\eta$.

Next, the correlation $\left<\phi'^2\right>$ shall be calculated. Coming from the bosonic theory of Luttinger liquids,\cite{giamarchi_book} one would be tempted to calculate it directly from the bMF Hamiltonian (\ref{toy_bMF}). This leads to $\left<\phi'^2\right>^{(0)} = K^*/2\eta^2$ and it will turn out that this way of calculating the correlation function is not correct: $K^* = {\tau^*}^{-1/2}$ diverges at the critical point, and so does $\left<\phi'^2\right>^{(0)}$. Thus, with this naive average, Eq. (\ref{toy_sc}) becomes ${\tau^*}^{\frac32} = \tau {\tau^*}^{\frac12} + 3\alpha$ and does not yield a proper solution with $\tau^*=0$, i.e., the system would never be able to reach its critical point. The reason that $\left<\phi'^2\right>^{(0)}$ is not the correct average is that at the critical point it diverges trivially in the sense that the Hamiltonian term governing the field $\phi$ becomes zero for $\tau^*=0$ and the Hamiltonian is independent of $\phi$.

If the dominant term controlling the field $\phi$ vanishes, previously subdominant terms become dominant and must be taken into account. The question is now, which out of the many terms in Eqs. (\ref{fws0_ham_bos}), (\ref{bos_fw_1}), and (\ref{bosonic_backscattering_red}) is the relevant one. Following the analogy to a Ginzburg-Landau theory, the quadratic second derivative term in Eq. (\ref{bos_fw_1}) is one candidate
\begin{equation}
\mathcal H^{(1)} = \beta \phi''^2.
\end{equation}
The average $\left<\phi'^2\right>$, calculated with $\mathcal H^{(0)}_{\rm bMF} + \mathcal H^{(1)}$ reads
\begin{equation}
\left<\phi'^2\right> = \frac14 \int \D k e^{-\eta|k|} \frac{k^2}{\sqrt{\tau^* k^2+ \beta k^4}}
\end{equation}
and is finite for $\tau^*=0$. The position of the critical point $\tau^*=0$ becomes
\begin{equation}
\tau = -\frac{3\alpha}{\eta\sqrt{\beta}}.\label{second_der_critical_point}
\end{equation}

The sine-Gordon Hamiltonian is also able to control the fluctuations of $\phi$ at the critical point. However, Hamiltonians involving trigonometric dependencies on a field are difficult to treat. In this work I only give a very simplified account of the sine-Gordon Hamiltonian by sketching one possible scenario based on the assumption that the conventional RG treatment of sine-Gordon terms in spinful Luttinger liquids\cite{giamarchi_book} extends to the present model.

Consider a bosonic Hamilton density
\begin{equation}
\mathcal H = u^* K^* \theta'^2 + \frac{u^*}{K^*}\phi'^2 + \frac{g_{1\perp}}{(2\pi\eta)^2} \cos(\sqrt8 \phi),
\end{equation}
where $\phi$ and $\theta$ are the spin sector fields of a spinful Luttinger liquid. $g_{1\perp}$ is the coupling constant of the backscattering processes. Note that in the presence of SU(2) symmetry there is a definite relation between $K^*$ and $g_{1\perp}$. The interaction $\cos(\sqrt 8 \phi)$ can be treated by a renormalization group analysis, within which $K^*\rightarrow1$ and $g_{1\perp} \rightarrow 0$. It is important to note that only for $K^*\rightarrow1$ the spin-spin correlation functions are independent of the direction of the spin-quantization axis chosen. In other words, the RG flow brings the system back to SU(2) symmetry. This calculation is perturbative in the cos term and thus it is only valid for sufficiently small $g_{1\perp}$. However, with the help of the SU(2) symmetry it may be argued that the validity of the RG idea can be extended to stronger interactions: only for $K^*\rightarrow1$ the theory describes an SU(2) invariant system and thus a system, in which this symmetry is inherent for all interaction strengths, should be correctly described by $K^*=1$ and $g_{1\perp}=0$, independently of the initial interaction parameters.

For the correlation function $\left<\phi'^2\right> = K^*/2\eta^2$ from the Hamiltonian $u^* K^* \theta'^2 + u^* \phi'^2/K^*$, this means with the sine-Gordon renormalized $K^*=1$ that $\left<\phi'^2\right>^{(\rm SG)}=1/2\eta^2$, which gives a critical point
\begin{equation}
\tau = -\frac{3\alpha}{\eta^2} \label{sg_critical_point}.
\end{equation}
Note, however, that Eq. (\ref{sg_critical_point}) is by no means the result of a controlled calculation. It should be interpreted as a sketch of a possible scenario for the critical behavior of weak edge magnetism.

Eqs. (\ref{second_der_critical_point}) and (\ref{sg_critical_point}) display two different possible scenarios for the critical behavior of edge magnetism. At this stage, however, it is not clear which scenario is the relevant one or if the fluctuations at the critical point are controlled by the combined action of the second derivative and the sign-Gordon term. Moreover, it is not clear which term in the Hamiltonian controls the critical properties at the transition. These questions are beyond the scope of this paper and are to be addressed in subsequent works.

\section{Summary and Discussion\label{sect_discussion}}

An interacting bosonic field theory has been derived from the fermionic model of weak edge magnetism. In this derivation it is crucial to properly account for the momentum dependence of the effective interaction vertex function, which finally leads to the bosonic interactions. This makes the bosonization mapping used here different from the conventional bosonization (see, e.g., Ref. \onlinecite{giamarchi_book}), where it is usually assumed that the interaction processes $g_{1,2,3,4}$ have an approximately momentum-independent strength. 

One of the most striking features of the bosonic theory derived in this work is the extended regime of validity. The usual theory of Luttinger liquids, applied naively to the edge states, breaks down for interactions $U$ that are strong enough so that $U/\pi v_F\geq 1$. But only above this bound, the physics becomes nontrivial by developing a spin polarization. Formally the breakdown of the naive theory happens because the bosonic Hamiltonian is not bounded from below in this case. For edge states, the momentum-dependence of the interactions leads to $\phi_s'^4$ terms with a positive prefactor. These terms ensure that the bosonic Hamiltonian is bounded from below and thus restore the validity of the theory for $U/\pi v_F>1$.

It is instructive to shed some light on the physical picture behind this mechanism. For a similar argumentation which is more targeted on an exact diagonalization analysis of edge magnetism, see Ref. \onlinecite{ed_luitz_2011}. Electron-electron interactions soften the Fermi level at zero temperature by exciting particle hole pairs to higher energies. For weak interactions, the kinetic energy penalty for these particle hole excitations is not overcompensated by the interaction energy gain. For sufficiently strong and momentum-independent interactions, however, overcompensation sets in and the electrons are excited to higher and higher energies by the interaction. Usually in one dimension, this process is not stopped until the the excitations hit the band edge. Thus, the band edge becomes important in this scenario, since it ultimately stops the flow of the system to high energy excitations. In bosonization, however, only effects effects in which the band edges play no role can be described properly. This means that bosonization breaks down as the particle-hole excitations hit the band edges. For momentum-independent interaction this happens exactly at the Stoner point (see also Ref. \onlinecite{ed_luitz_2011}). The strong momentum dependence of the interaction vertex of edge states, which makes the interaction effectively weaker at high energies (see Fig. \ref{fig_fermionic_model}), stops the excitations before they hit the band edge. This is how the non-standard effective interaction vertex of edge states saves the bosonization from breaking down in the regime of weak edge magnetism. Thus, the special feature of strongly momentum-dependent interactions of edge states is crucial for the applicability of bosonization to one-dimensional magnetism.

The bosonic theory, consisting essentially of the three Hamiltonians (\ref{fws0_ham_bos}), (\ref{bos_fw_1}), and (\ref{bosonic_backscattering_red}), seems to be extremely complicated at first glance. However, it paves the way to a separation of important and unimportant terms. The universal critical behavior is expected to depend only on a small number of qualitatively important terms in the bosonic field theory, and a goal of future investigations should be to identify those terms and to extract the critical properties of the ferromagnetic transition between the Luttinger liquid and the weak ferromagnetism. For instance, it was demonstrated in this work on a mean-field level that the terms $\tau\phi_s'^2 + \phi_s'^4$ are not sufficient to describe the critical point. One rather needs to include a term in the Hamiltonian which controls the spatial fluctuations of the magnetization at the critical point. On the other hand, the term $\theta_c'^4$ in Eq. (\ref{fws0_ham_bos}), for example, is expected not to affect the magnetic properties qualitatively, since a positive $\theta_c'^2$ term already limits the fluctuations of $\theta_c'$.

A thorough analysis of the singularities of this theory will give rise to the identification the qualitatively important terms. From the qualitative analysis of the bosonic field theory presented in this work, one may already state the preliminary expectation that the critical properties of the tunable edge magnetism is described a theory of the form
\begin{equation}
\mathcal H_{\rm crit.} = \theta_s'^2 + \tau \phi_s'^2 + \phi_s'^4 + \phi_s''^2 + g \cos\sqrt8\phi_s \label{critical_hamiltonian}
\end{equation}
where the first term controls the quantum fluctuations (remember that $\phi_s$ and $\theta_s$ are conjugate fields), the second and third terms are the Landau function part of the theory, and the last two terms control the field fluctuations at the critical point. On the basis of this 1+1 dimensional classical field theory, the effect of other terms on the critical behavior may be studied systematically, but it is expected that the essential physics at the transition is captured by $\mathcal H_{\rm crit.}$.

Note that the fields entering Eq. (\ref{critical_hamiltonian}) are real and $\theta_s$ enters only quadratically. Thus, $\theta_s$ can be integrated out exactly and the critical theory can be formulated as a classical field theory with a real action. This enables the investigation by Monte-Carlo methods, since theories which can be formulated with real actions have no sign problem.

It is furthermore important to note that the role of SU(2) invariance it is not clear at this stage. In the traditional bosonization approach to Luttinger liquids,\cite{giamarchi_book} the sine-Gordon Hamiltonian, describing the backscattering processes, is responsible for restoring SU(2) invariance in the correlation functions. A perturbative renormalization group treatment leads to a non-interacting Hamiltonian with renormalized parameters ($K_s^*=1$) in the spin sector. The question in which way these arguments hold in the bosonic field theory of edge magnetism, derived here, is beyond the scope of this work.

\acknowledgments
I would like to acknowledge interesting and enlightening discussions with F. F. Assaad, B. Braunecker, D. Loss, and D. J. Luitz. This work was supported by the Swiss NSF and by the NCCR QSIT. 

\appendix

\section{Bosonization details\label{appendix_bosonization_details}}
A generalized bosonization technique for momentum-dependent electron-electron interactions of the form (\ref{def_h_fs}) or (\ref{eq_ham_backscattering}) is derived. In terms of the real space fields $\psi_{r\sigma}(x) = L^{-1/2} \sum_k e^{ikx} c_{kr\sigma}$, the forward scattering Hamiltonian reads
\begin{multline}
H^{\rm fs}_1 = U \sum_{rr'}\int \D x \\ \left[S^r(-\hat k) \psi^\dagger_{r\uparrow}\right] \left[S^r(\hat k) \psi_{r\uparrow}\right] \left[S^{r'}(-\hat k) \psi^\dagger_{r'\downarrow}\right] \left[S^{r'}(\hat k) \psi_{r'\downarrow}\right],
\end{multline}
where $S^r(k) = \sqrt{1-r\Gamma_1 k}$ and $\hat k = -i\partial_x$ is the momentum operator. $r=R,L$ labels right- and left-moving fermions, respectively. Here and henceforth, normal order with respect to the non-interacting ground state is implicitly assumed. $H^{\rm fs}_1$ is expanded in powers of the momentum operators, acting on the different fermionic fields $\psi_{r\sigma}$. The goal is then to translate the typical terms
\begin{equation}
\partial_{x_1}^k  \partial_{x_2}^l \partial_{x_3}^m \partial_{x_4}^n \psi^\dagger_{r\uparrow}(x_1)  \psi_{r\uparrow}(x_2)  \psi^\dagger_{r'\downarrow}(x_3) \psi_{r'\downarrow}(x_4), \label{general_interaction_form}
\end{equation}
of which the expanded $H_1^{\rm fs}$ is composed, to a bosonic language. The four different spatial coordinates $x_1,...,x_4$ are used in order to keep track of the correspondence between the differential operators and the fermion fields, and to be able to bosonize these terms by a point splitting method. In the end, the limit $x_1,x_2,x_3,x_4\rightarrow x$ must be taken. 

Since this work is concerned with Hubbard interactions only, the quartic Fermion terms always consist of two spin-up and two spin-down operators. This simplifies the analysis as the quadratic terms may be bosonized separately for each spin species. For the backscattering interaction [Eq. (\ref{eq_ham_backscattering})] the same idea as described above is used. However, the details of the procedure differ slightly compared to the forward scattering interaction. Note also that, as far as the bosonization is concerned, it is always assumed that the thermodynamic limit $L\rightarrow\infty$ is performed in the end, so that only the leading terms in $L^{-1}$ are to be kept.

\subsection{The bosonic fields}

Following Ref. \onlinecite{giamarchi_book}, the fermionic fields may be expressed in terms of exponentiated bosonic fields
\begin{equation}
\psi^\dagger_{r\sigma}(x) = \frac1{\sqrt L} e^{-i\gamma_{r\sigma}^\dagger(x)} e^{-i\gamma_{r\sigma}(x)},\label{bos_identity}
\end{equation}
where $r=R,L$ labels right- and left-moving Fermions, respectively, and $\sigma$ labels the spin. The Klein factors in Eq. (\ref{bos_identity}) have been dropped, since they appear only in the standard combinations for which it is well known that they are irrelevant.\cite{giamarchi_book} The bosonic (but non-Hermitian) fields $\gamma_{r\sigma}(x)$ are defined as
\begin{equation}
\gamma_{r\sigma}(x) = \frac{r \delta \hat N_{r\sigma}\pi}L x + \sum_{q>0} \sqrt{\frac{2\pi}{L q}} e^{irqx - \frac\eta2 q}b_{qr\sigma},\label{gamma_def}
\end{equation}
where $\delta \hat N_{r\sigma} = \sum_k \left[ c^\dagger_{kr\sigma} c_{kr\sigma} - \left< c^\dagger_{kr\sigma} c_{kr\sigma}\right>\right]$ is the total density of $r$-movers with spin $\sigma$, relative to the non-interacting ground state. The $k$-space boson operators, in terms of the original fermionic operators $c_{kr\sigma}$, read
\begin{equation}
b_{qr\sigma} = -i \sqrt{\frac{2\pi}{Lq}} \sum_k c^\dagger_{k-rq,r,\sigma} c_{kr\sigma}. \label{b_def}
\end{equation}
Note that $b_{qr\sigma}$ is only defined for $q>0$. On the basis of the fermionic commutation rules for $c_{kr\sigma}$, it can be shown\cite{giamarchi_book} that $\left[b_{qr\sigma},b^\dagger_{q'r'\sigma'}\right] = \delta_{qq'}\delta_{rr'}\delta_{\sigma\sigma'}$. At some points in the bosonization procedure, expressions need to be regularized. This is done in the standard way\cite{giamarchi_book} by introducing an exponential UV cutoff $e^{-\eta |q|/2}$ in the $k$-space sums. $\eta$ has the meaning of a microscopic length scale, such as the lattice constant, thus a number of order one. In most of what follows, $\eta$ can be regarded as small compared to typical distances. However, if $\eta$ needs to be specified explicitly, it should be chosen to be the inverse width of the reduced Brillouin zone, i.e. $\eta\simeq 6/\pi$. 

In the final bosonized expressions, only the real parts of the four different $\gamma$ fields appear, i.e.
\begin{equation}
\phi_{r\sigma}(x) =\frac12 (\gamma^\dagger_{r\sigma}(x) + \gamma_{r\sigma}(x)). \label{def_phi_rs}
\end{equation}
The spatial derivatives of $r\phi_{r\sigma}$ may be interpreted as a local density of $r$-moving fermions with spin $\sigma$, as is easily seen by substituting Eqs. (\ref{gamma_def}) and (\ref{b_def}) in Eq. (\ref{def_phi_rs}). The symmetric and antisymmetric combinations of left- and right-moving fermions give rise to the charge and current densities with spin $\sigma$
\begin{align}
\phi_\sigma(x) &= \sum_r (r\phi_{r\sigma}(x)) & 
\theta_\sigma(x) &= - \sum_r r(r\phi_{r\sigma}(x)).
\end{align}
Furthermore, the symmetric and antisymmetric combinations of opposite spins form the charge and spin sector basis
\begin{align}
\phi_c &= \frac1{\sqrt2}(\phi_\uparrow+\phi_\downarrow) & \phi_s &= \frac1{\sqrt2}(\phi_\uparrow - \phi_\downarrow) \\
\theta_c &= \frac1{\sqrt2}(\theta_\uparrow+\theta_\downarrow) & \theta_s &= \frac1{\sqrt2}(\theta_\uparrow - \theta_\downarrow) .
\end{align}

Note that the field $\gamma_{r\sigma}$ does not commute with its complex conjugate $\gamma_{r\sigma}^\dagger$. From Eq. (\ref{gamma_def}), the commutation rules of the $\gamma$ fields may be calculated
\begin{equation}
\left[\gamma_{r\sigma}(x),\gamma^\dagger_{r'\sigma'}(x')\right] = \delta_{rr'}\delta_{\sigma\sigma'} \log \frac{L}{2\pi(\eta-ir(x-x'))}.\label{gamma_commutator}
\end{equation}
The subleading terms in $L^{-1}$ have been dropped.

\subsection{Forward scattering}
As explained above, the two factors for each spin species in Eq. (\ref{general_interaction_form}) may be treated separately. The form of these two factors is the same for both, up-spin and down-spin. The general form of the terms to be bosonized is
\begin{equation}
F[n,m] = \left(i\partial_x\right)^n\left(-i\partial_{x'}\right)^m : \psi^\dagger_{r\sigma}(x) \psi_{r\sigma}(x') :
\end{equation}
$F[n,m]$ contains all factors corresponding to spin $\sigma$ in Eq. (\ref{general_interaction_form}). The normal order $:A: = A - \left<A\right>$ of an operator $A$ which is a quadratic form of fermion operators, with respect to the non-interacting ground state is needed to regularize the theory.\cite{giamarchi_book} The normal order arises directly from the mean-field treatment of the direct model.\cite{tem_schmidt_loss_2010} From Eq. (\ref{bos_identity}), one finds
\begin{multline}
\psi^\dagger_{r\sigma}(x) \psi_{r\sigma}(x') = \\\frac{\exp i(\gamma^\dagger_{r\sigma}(x')-\gamma^\dagger_{r\sigma}(x)) \exp i(\gamma_{r\sigma}(x')-\gamma_{r\sigma}(x))}{2\pi(\eta+ir(x'-x))},\label{fermions}
\end{multline}
where the two exponential factors in the middle have been commuted and thereby gave rise to the factor $\exp \left[\gamma_{r\sigma}(x),\gamma_{r\sigma}^\dagger(x')\right] = L/2\pi(\eta+ir(x'-x))$ [see Eq. (\ref{gamma_commutator})]. After having applied the spatial derivatives $(i\partial_x)^n(-i\partial_{x'})^m$ to the right hand side of Eq. (\ref{fermions}), the expression is expanded into a Taylor series in $\delta x = x'-x$. The terms of order $\delta x$ and higher vanish, the terms of order $\delta x^0$ are the bosonic forms of $F[n,m]$ and the terms of order $\delta x^{-i},\sms i\geq 1$ are constant (and diverging) real numbers which are removed by the normal order.

The terms of zeroth and first order in the derivatives, resulting from the procedure described above, are
\begin{align}
F[0,0] &= \frac r\pi \phi'_{r\sigma} \\
F[0,1] &=\frac r\pi (\phi'_{r\sigma})^2  - i\partial_x  \frac {r}{2\pi}\phi'_{r\sigma}\\
F[1,0] &=\frac r\pi (\phi'_{r\sigma})^2  + i\partial_x  \frac {r}{2\pi}\phi'_{r\sigma}
\end{align}
As expected, $F[0,0]$ is a local density term. The non-Hermitian terms, appearing in $F[0,1]$ and $F[1,0]$, are canceled in the final Hamiltonian. Note also that $F[m,n-m]$ for different $m$ but $n$ fixed differ only by a total derivative (TD). This is enforced by the Hermiticity of the momentum operator $-i\partial_x$. For a Hamiltonian of the form $\int\D x F[n,m]$, these TD terms may be removed, as they only give rise to irrelevant boundary terms. However, since $F[n,m]$ enters $H^{\rm fs}_1$ quadratically, the TD terms must be kept in principle. It turns out, however, that they do not give essential contributions to the physics. Therefore, we partition $F[m,n-m]$ into
\begin{equation}
F[m,n-m] = F^{(0)}[n] + \partial_x F^{(1)}[m,n-m]
\end{equation}
and discuss only the terms $F^{(0)}$ for now.

Each spin ($\sigma=\uparrow,\downarrow$) contributes a factor
\begin{widetext}
\begin{multline}
\sum_r \left[S(ir\partial_x) \psi^\dagger_r(x)\right] \left[S(-ir\partial_{x'}) \psi^\dagger_r(x')\right] = \sum_r  \sum_n (r\Gamma_1)^n \sum_{m=0}^n c_m c_{n-m} F[m,n-m]  \\ =  \sum_r  \sum_n (r\Gamma_1)^n F^{(0)}[n] \sum_{m=0}^n c_m c_{n-m}  + \partial_x ...
\end{multline}
\end{widetext}
to the integrand of the Hamiltonian $H_1^{\rm fs}$. $c_n=\frac{(2n-3)!!}{2^n n!}$ is the prefactor of the $n$th order in the Taylor series of $\sqrt{1-x}$. The last expression indicates the TD terms which are dropped from in the remainder of this section. For $n>1$, the convolution of the $c_n$ gives zero, so that
\begin{equation}
H^{{\rm fs},(0)}_1 = \frac U{\pi^2} \sum_{r,r'} \int \D x \left[r\phi'_{r\uparrow} - \Gamma_1 \phi'^2_{r\uparrow}\right] \left[r\phi'_{r\downarrow} - \Gamma_1 \phi'^2_{r\downarrow}\right].\label{fsham_rsigma}
\end{equation}
Transforming Eq. (\ref{fsham_rsigma}) to the spin and charge sector basis, we arrive at Eq. (\ref{fws0_ham_bos}).

\section{Derivation of the bosonic mean-field equations\label{appendix_bMF}}

We use Feynman's variational principle in order to find the best approximation of an interacting theory by a non-interacting theory. More explicitly, the goal is to find a quadratic variational action $S_v$ for a given action $S$, so that the inequality
\begin{multline}
\beta F = -\ln Z = -\ln \int \mathcal D \Phi e^{-S_v} e^{-(S-S_v)} \\ =\beta F_v - \ln \left<e^{-(S-S_v)}\right>_v \leq \beta F_v + \left<S-S_v\right>_v\label{var_principle}
\end{multline}
is satisfied best. In other words, the goal is to find an action for which the right hand side of Eq. (\ref{var_principle}) is minimal. Assuming that $S=S_0 + S_1$ consists of a free part and an interacting part, we make the most general ansatz
\begin{equation}
S_v = S_0 + \frac1{2\pi\beta L} \sum_{k,\omega} \Phi^\dagger_{k,\omega} \cdot \ve M(k,\omega) \cdot \Phi_{k,\omega}
\end{equation}
where $\Phi_{k,\omega}$ is a vector of spatial derivatives of fields in $(k,\omega)$ space. The matrix elements of $\ve M(k,\omega)$ are to be determined by
\begin{equation}
\delta\left[\beta F_v + \left<S-S_v\right>_v \right]=0.
\end{equation}
As usual, this method gives rise to the following replacement rule for quartic interactions
\begin{align}
\phi^\mu\phi^\nu\phi^\eta\phi^\tau \overset{MF}\longrightarrow &\sms \contraction{}{\phi}{{}^\mu}{\phi}
\phi^\mu\phi^\nu\phi^\eta\phi^\tau +
\contraction{}{\phi}{{}^\mu\phi^\nu}{\phi}
\phi^\mu\phi^\nu\phi^\eta\phi^\tau +
\contraction{}{\phi}{{}^\mu\phi^\nu\phi^\eta}{\phi}
\phi^\mu\phi^\nu\phi^\eta\phi^\tau \nonumber\\&+
\contraction{\phi^\mu}{\phi}{{}^\nu}{\phi}
\phi^\mu\phi^\nu\phi^\eta\phi^\tau +
\contraction{\phi^\mu}{\phi}{{}^\nu\phi^\eta}{\phi}
\phi^\mu\phi^\nu\phi^\eta\phi^\tau +
\contraction{\phi^\mu\phi^\nu}{\phi}{{}^\eta}{\phi}
\phi^\mu\phi^\nu\phi^\eta\phi^\tau,\label{replacement_rule}
\end{align}
where the fields $\phi^\mu$ are components of the field vector $\Phi$ and the contractions can be calculated by quadratic field averages
\begin{equation}
\contraction{}{\phi}{{}^\mu}{\phi}\phi^\mu\phi^\nu = \left<\phi^\mu(x,\tau)\phi^\nu(x,\tau)\right>_v.
\end{equation}
In the present case, the quartic interaction term is composed of $\phi'$, the spatial derivative of the bosonic field which describes the local spin density. Thus, all the six contractions are equal and one obtains
\begin{equation}
\alpha \phi'^4 \overset{MF}\longrightarrow 6\alpha \langle\phi'^2\rangle\phi^2.
\end{equation}
Note that Eq. (\ref{replacement_rule}) in its full generality can also be used for a mean-field treatment of the more complete model (\ref{fws0_ham_bos}). It is, however, not obvious how the sine-Gordon term may be treated within this approach. All these issues are beyond the scope of this work and are to be investigated in future studies.

\bibliography{tem_bib}

\end{document}